\documentclass[12pt]{article}
\usepackage{psfig}
\usepackage{cite}
\begin{document}

\title{Transition temperatures and contact angles in the sequential-wetting 
scenario of $n$-alkanes on (salt) water}
\begin{small}
\author{Volker C.\ Weiss\thanks{Corresponding author. Tel.: +32 16 327 220,
        Fax: +32 16 327 983. \newline E-mail: 
        volker.weiss@fys.kuleuven.ac.be (V.C.\ Weiss)} \, and 
        Joseph O.\ Indekeu\\[0.5cm]
        Laboratorium voor Vaste-Stof\/fysica en Magnetisme,\\
        Katholieke Universiteit Leuven, B-3001 Leuven, Belgium\\[0.3cm]}
\date{\today}
\maketitle
\end{small}

\begin{abstract}
\noindent Alkanes of medium chain length show an unusual wetting
behavior on (salt) water, exhibiting a sequence of two changes in
the wetting state. When deposited on the
water surface at low temperatures, the liquid alkane forms discrete
droplets that are interconnected only by a molecularly thin film.
On increase of the temperature, there occurs a sudden jump of the film
thickness and, at this first-order transition, a mesoscopically thick
layer of liquid alkane is formed. 
Heating the system further leads to a divergence of the film thickness
in a continuous manner.
While the long-range forces between substrate and adsorbate are responsible
for the critical transition, which occurs at the temperature at which
the Hamaker constant changes sign, it is primarily the short-range
components of the forces
that bring about the first-order transition. We calculate the Hamaker 
constant of the system and show 
how, within a modified Cahn theory, accurate predictions of the first-order
transition temperatures can be obtained for $n$-alkanes (pentane and hexane) 
on water and even on brine. Furthermore, we study the variation of the contact
angle as a function of temperature.   
\end{abstract}
\vspace{1cm}
\newpage

\section{Introduction}
In recent experiments, $n$-alkanes of medium chain length (pentane,
hexane, and heptane) have been observed to undergo two, instead of
just one, changes of the wetting state when they are deposited on
(salt) water and the temperature is raised subsequently \cite{shahidzadeh,
bertrand,bertrandth}. At sufficiently
low temperature, the deposited liquid alkane, which is at coexistence
with its vapor, forms discrete droplets on the surface of the substrate.
These droplets make a non-zero contact angle with the substrate surface
and, for volatile substances like alkanes of relatively low molecular
weight, are interconnected only by a molecularly thin film. This state
is referred to as partial wetting. At high enough temperatures, on the 
other hand, there will be a macroscopically (in principle, infinitely)
thick film of liquid alkane covering the (salt) water surface completely.
The contact angle in this so-called complete-wetting state is exactly zero.
The transition between these two states can, in general, occur in two
qualitatively different fashions: at a first-order transition, one observes
an abrupt change of the film thickness from microscopic to macroscopic
values; likewise, the contact angle $\theta$, which becomes smaller with 
increasing temperature, vanishes and, in doing so, its derivative with 
respect to temperature displays a square-root singularity of the form 
$d \theta / d T \sim \left( T_w - T\right)^{-1/2}$, where $T_w$ denotes
the wetting-transition temperature. In the second and much rarer case,
the film thickness diverges continuously on approach of the wetting 
temperature \cite{ragil,ross}. This type of wetting transition is, therefore, 
referred to as being continuous or `critical'. Although predicted earlier
from theoretical considerations \cite{critwet,dietrich}, it had escaped 
experimental observation until recently. 
For an $n$-nonane/methanol mixture, for which the wetting transition
temperature is very close to the consolute point, so that bulk fluctuations
become important, a logarithmic divergence of the film thickness on
approaching $T_w$ has been observed \cite{ross}, as well as a vanishing of 
the contact angle according to the power law $1 - \cos{\theta} \sim 
(T_w - T)^{2 - \alpha_s}$, with $\alpha_s$ being the surface specific
heat exponent and having a numerical value of $-0.2 \pm 0.3$, which
is compatible with short-range critical wetting ($\alpha_s = 0$,
implying $\theta \sim \left( T_w - T \right)$), but 
not with first-order wetting ($\alpha_s = 1$).    
In the case of long-range critical wetting, in contrast, which was
observed first for pentane on water,  one finds a power-law divergence of
the film thickness of the form $l \sim \left( T_w - T \right)^{-1}$
\cite{ragil} (and the contact angle vanishes according to $\theta \sim 
\left( T_w - T \right)^{3/2}$, corresponding to a surface specific heat 
exponent of $\alpha_s = -1$). This 
behavior follows from the leading
terms of the tails of the long-range forces between substrate and
adsorbate \cite{dietrich}. 

By now, it has become apparent that such a long-range critical
wetting transition is a generic feature of all $n$-alkanes of medium
chain length on water or brine, respectively
\cite{bertrand,bertrandth,bonnross}.
Salt was originally added to the aqueous substrate to decrease the
wetting temperature in such a way that the transition occurs in the
experimentally accessible range of $0-80^{\circ}$C, i.e.\ for experimental
convenience \cite{shahidzadeh}, but the cases in which brine, rather than
pure water, is the substrate seem to be more relevant in industrial and
environmental situations of practical interest, for example in
secondary oil recovery \cite{bertrandth,josephnato,broseta}. 

The occurrence of long-range critical wetting in systems of alkane
on (salt) water gives rise to another peculiar feature: the transition
from partial to complete wetting takes place in two steps instead of
just one! \cite{shahidzadeh} As mentioned above, the true wetting transition 
to complete
wetting is continuous and brought about by long-range forces between
substrate and adsorbate, which change their nature from inhibiting
complete wetting to supporting it at this temperature, $T_{w,c}$. This
critical transition, however, is preceded by a first-order transition
at a lower temperature, $T_{w,1}$, at which the film thickness shows an abrupt
increase from being molecularly thin to a mesoscopic value (6--15 nm)
\cite{shahidzadeh,bertrand}.
Accordingly, the temperature-derivative of the contact angle shows a
finite discontinuity (while $\theta$ itself is a continuous function
of $T$) at $T_{w,1}$ \cite{weiss}, which would be (very close to) the 
first-order {\em wetting}\/
transition temperature if the long-range forces did not oppose complete
wetting, i.e.\ the formation of a macroscopic wetting layer, at this
lower temperature. The occurrence of such a transition, the 
first-order
nature of which is corroborated by the experimentally observed hysteresis
in the film thickness \cite{shahidzadeh}, is attributable to the short-range
parts of the forces, as taken into account in Cahn's original theory of 
wetting transitions \cite{ragil,indekeu,cahn}.

For the intermediate state, in which a film of mesoscopic thickness is
present, the term `frustrated-complete wetting' has been coined
\cite{josephnato}. For
a typical film thickness of 100 {\AA}, the contact angle of the droplets
that sit on top of this layer is very small ($< 1^{\circ}$), indicating
that this state is energetically very close to complete wetting
\cite{bertrand,weiss}. In this
state, the efficiency of oil recovery by drainage from a model reservoir 
is also similar to that from a system in the complete wetting state
\cite{bertrandth}. On grounds of dynamic arguments, however, one
expects considerable differences to the flow in a system with a truly 
macroscopic wetting layer \cite{broseta2}.
Therefore, accurate knowledge of {\em both}\/ transition temperatures
is highly desirable.

In the next section, we will outline the main features of a complete and 
consistent mean-field theory of the wetting behavior of $n$-alkanes
of medium chain length (pentane and hexane) on pure water and on brine,
respectively, for various concentrations of salt. In section 3, we
will compare the results of our theory to experimental findings and 
predict the transition temperatures and contact angles for systems
for which these properties have not been measured yet. Section 4 concludes
the paper, assesses the current state of the theory and experiments
on the sequential wetting of $n$-alkanes on aqueous substrates and
briefly outlines plans for future investigations.
 
\section{Methodology}
The wetting state, in which a system consisting of three phases
(the substrate ($s$), which may be solid or liquid, the liquid ($l$) 
phase of the adsorbate and vapor ($v$)) 
is present, is determined by the interfacial
tensions between the three pairs of them, which we denote by $\gamma_{sl}$,
$\gamma_{sv}$ and $\gamma_{lv}$. According to Young's law, the contact
angle is given by $\gamma_{sv} - \gamma_{sl} = \gamma_{lv} \cos{\theta}$.
In the partial-wetting state, $\gamma_{sv} < \gamma_{sl} + \gamma_{lv}$
and, therefore, $\theta > 0$, while for complete wetting, the $sv$-interface
is replaced an $sl$-interface and an $lv$-interface and, accordingly,
Antonow's rule $\gamma_{sv} = \gamma_{sl} + \gamma_{lv}$ holds. The
contact angle is then zero. In the special case of sequential wetting,
the ultimate transition to complete wetting, which occurs at $T_{w,c}$,
is determined by the location at which the Hamaker constant changes sign,
i.e.\ where the leading term of the algebraic tails of the long-range forces
starts to favor having a thick wetting layer on top of the substrate,
instead of opposing the formation of such a layer, which keeps the
system in the frustrated complete wetting state for $T_{w,1} \le T \le 
T_{w,c}$. The first-order transition between a thin and a mesoscopically
thick film, however, is not a true wetting transition. Its location is
determined mainly by short-range forces \cite{indekeu,dobbs}, but the 
long-range tails, which are treated perturbatively in our approach, make a 
quantitative difference \cite{weiss3}. $T_{w,1}$ is found by comparing the 
interfacial free energies of 
the situations corresponding to having only a molecularly thin film present
in the density profile of an $sv$-interface and to having a mesoscopically
thick film -- only one of these two will be stable, the other just
metastable, except for right at $T_{w,1}$, where both are equal in
free energy per unit area (coexistence of thin and thick film).

To calculate the interfacial tensions, we employ a modified Cahn
theory augmented for algebraically decaying long-range forces, which
are treated in a perturbative fashion. The free-energy functional,
which is to be minimized with respect to the density profile of the
adsorbate, $\rho(z)$, reads \cite{weiss3}:
\begin{eqnarray} 
   \gamma[\rho] & = & \gamma_0 + \phi(\rho_0) + \int\limits_{\Delta
                  z}^{\infty}
                  \left\{ \Delta f(\rho, \rho_{bulk}) + \frac{c}{2}
                  \left( \frac{d \rho}{d z} \right)^2 \right\} dz \nonumber
                  \\
                & & +\left\{ \Delta f(\rho_0, \rho_{bulk}) + \frac{c}{2}
                  \left[(\rho_0 - \rho_1) / \Delta z\right]^2\right\}
                  \Delta z \nonumber \\
                & & - \int\limits_{z_c}^{\infty} \left( \frac{a_3}
                  {z^3} + \frac{a_4}{z^4}\right) \rho(z) dz
                  - \int\limits_{z_c}^{\infty} \left(\frac{a_3'} {(z +
                  \delta)^3} + \frac{a_4'}{(z + \delta)^4}\right) \rho(z) dz
                  \nonumber \\
                & & - \frac{a_3'}{\delta^3} \rho_0 \Delta z.
                  \label{intfactensalt}
\end{eqnarray}
The first three terms are the ingredients of standard Cahn theory (with
$\Delta z = 0$) \cite{cahn}: the first one describes the surface tension of 
the substrate against vacuum; it is irrelevant to the wetting properties
and no value for it needs to be specified. In the second term, $\phi$
denotes the so-called contact energy, which describes the short-range
interaction between the substrate and the adsorbate; therefore, it only
depends on the density of the adsorbate right at the substrate surface,
$\rho(0) = \rho_0$. The $z$-axis is chosen perpendicular to the substrate
surface and only the half space $z \ge 0$, in which the adsorbate is
present, is considered explicitly. $\phi\left(\rho_0\right)$ is usually
represented in a quadratic form: $\phi\left(\rho_0\right) = - h_1 \rho_0
- g \rho_0^2/2$; the coefficients $h_1$ and $g$ can be deduced from
measurements of the surface pressure \cite{dobbs,ragil2}. The third term 
concerns the
cohesive properties of the adsorbate, like in the classical theory
of the liquid--vapor interfacial tension of van der Waals \cite{rowlinson}: 
$\Delta 
f(\rho,\rho_{bulk})$, to be calculated from the Peng--Robinson equation
of state in this case, measures the excess free energy of 
locally having
a density $\rho(z)$ over having either of the bulk densities $\rho_{bulk}$;
the second term in the integral is the square-gradient term, which
penalizes inhomogeneities of the density. The coefficient $c$ is the
influence parameter and is obtained by matching the computed liquid--vapor
interfacial tensions to experimental results \cite{carey}. In the modified 
Cahn theory,
these two terms are to be integrated over from $\Delta z$ to infinity,
instead of integrating from zero to infinity as in standard Cahn theory
\cite{cahn}. 
The first layer of adsorbed alkane molecules, which is of molecular
thickness ($\Delta z = \sigma$, the diameter of an adsorbate
molecule) is treated  in a lattice-gas approach, ensuring that
states of low adsorption are properly accounted for \cite{dobbs}. This 
discretized
version of Cahn theory yields the next term in Eq.~(\ref{intfactensalt}) --
its ingredients are otherwise the same as in the continuum version.
For brevity, the density at a distance of one molecular diameter
from the substrate surface is denoted by $\rho_1 = \rho(\Delta z)$.

The last three terms on the right-hand side of Eq.~(\ref{intfactensalt}) 
incorporate the
long-range interaction into the free-energy functional. The underlying
picture of the system in the computation of the long-range contribution
is a five-layer structure \cite{weiss3}: even if the substrate is brine, 
there is a
thin layer of pure water (the so-called depletion layer) at the brine/alkane
interface \cite{onsager,levin2}, so that it is always pure water that is in 
direct contact with the
adsorbed alkane. A third (conceptual)
medium is formed by the dense liquid layer of adsorbed alkane molecules
which are in direct contact with the aqueous substrate
\cite{bertrandth,bertrandepl}. Moving away from
the substrate surface, this layer is followed by the bulk liquid alkane
and, ultimately, by the vapor phase. While the brine and the vapor
phases are semi-infinite, the pure-water layer is of thickness $\delta$ (to
be identified with the hydration radius of an ion \cite{levin2}),
the first layer of adsorbed dense liquid alkane has a spatial extent
of one molecular diameter $\sigma = \Delta z$, and the film thickness of 
liquid alkane is denoted by $l$.

For pure water as the substrate, the last two terms
in Eq.~(\ref{intfactensalt}) vanish. The first of the three terms
of the long-range field decribes the interaction between the water layer
and the adsorbate; the cutoff $z_c$ is numerically very close to
$1.5 \sigma$, the distance of closest approach of adsorbate particles
that are not part of the first layer. The second integral concerns
the long-range interactions between brine and the adsorbate across
the intervening water layer of thickness $\delta$, which is reflected
by having $(z + \delta)$ instead of just $z$ in the denominators of
the first couple of leading terms taken into account here.  The last
term in Eq.~(\ref{intfactensalt}) contains the remaining contribution
which arises from the interaction of brine with the first layer of
adsorbed alkane molecules -- this interaction is not included in
the preceding term since $z_c > \sigma$.

The amplitudes of the tails of the long-range field, $a_3$, $a_4$,
$a'_3$ and $a'_4$ are calculated from a five-layer
Dzyaloshinskii--Lifshitz--Pitaevskii theory that
invokes Israelachvili's simplifications to the dielectric spectrum
\cite{israel},
which allows one to obtain the long-range field from the static
dielectric permittivities, $\varepsilon_j(0)$, the indices of refraction,
$n_j$, and a common characteristic UV absorption frequency of the different
media. The explicit formulas for the amplitudes can
be found in the more detailed article \cite{weiss3}.

The current theory offers a complete and consistent description of the
wetting behavior of $n$-alkanes on water or brine, as we will show in the
next section. 

\section{Results and discussion}
We employ the theory outlined in the preceding section 
to calculate the transition temperatures as a function of the salinity
of the substrate and contact angles as a function of temperature in the 
sequential wetting of $n$-alkanes, namely pentane and hexane, on
salt water. Figure 1 shows the calculated first-order and critical transition
temperatures for pentane and hexane, respectively, on brine in the
concentration range $0 \le c_{\rm NaCl} \le 2.5$ mol/L\@. The experimental
points for the respective transition temperatures for hexane on brine 
\cite{shahidzadeh} and for pentane on water \cite{ragil,bertrand} are 
indicated by open symbols. The overall
agreement is very good, indicating that the theory incorporates the 
most important effects that lead to sequential wetting even
quantitatively. Note that the first-order transition temperatures for
pentane on brine are below the freezing point of the substrate for
$c_{\rm NaCl} > 2.1$ mol/L\@. Furthermore, it is seen that, within 
the theory, the line
of first-order transition temperatures is almost parallel to the line
of critical-transition temperatures for pentane
on brine, too, just as it was observed experimentally for hexane on
brine \cite{shahidzadeh}. 
An experimental check of this prediction is desirable.

\begin{figure}
\centerline{\psfig{figure=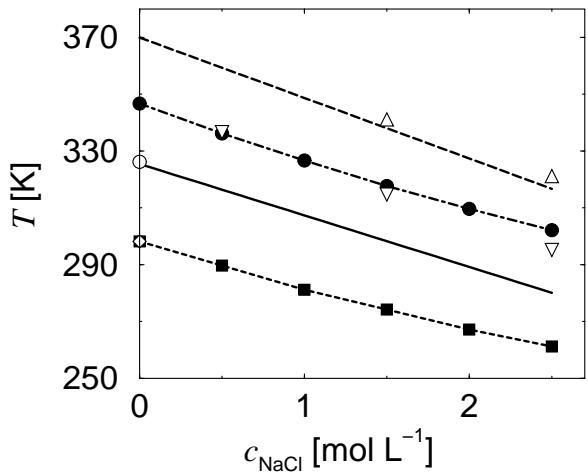,width=7cm,angle=-90}}
\caption{First-order and critical transition temperatures
for pentane and hexane, respectively, on brine as a function of salinity. The 
open symbols correspond to experimentally 
determined temperatures; lines and filled symbols mark calculated transition
temperatures: critical wetting for hexane on brine (open triangles up, 
long-dashed line), first-order transition for hexane on brine (open triangles 
down, filled circles and dot-dashed line), critical wetting
for pentane on water/brine (open circle, solid line), first-order transition 
for pentane on water/brine (open diamond, filled squares, dashed line). 
The calculated values were obtained using $\delta = 1.9$ \AA\@. }
\end{figure}

To test the validity of the more or less arbitrary division of the
forces acting between substrate and adsorbate into short-range and long-range
parts, and of the perturbative treatment of the long-range parts, it might 
be useful to take a closer look at the contact angles,
in particular in the frustrated-complete wetting state. In this latter
state, the contact angle is very small ($< 1^{\circ}$), indicating 
that the relatively large film thickness of $l \ge 60$ {\AA} is sufficient
to decouple the two interfaces ($sl$ and $lv$) almost completely \cite{weiss}. 
In the context of the present theory, the contact angle can, asymptotically, 
be evaluated from
the long-range part alone \cite{bertrandth}. To accomplish this, one uses the 
fact that the contact angle is related to the spreading coefficient through 
$S = \gamma_{lv} (\cos{\theta} - 1)$. We consider the free-energy difference 
between the $sv$-interfaces in the complete wetting ($cw$) state and the
frustrated-complete wetting ($fcw$) state, the origin of which lies completely 
in the range of distances between $l$ and infinity from the substrate. Using
$S_{cw} = 0$ and the fact that $\rho(z)$ is practically equal to $\rho_l$
for $cw$, while it is almost zero (the vapor phase is very dilute) in 
this range for $fcw$, we obtain for the spreading coefficient in the $fcw$ 
state:
\begin{equation} \label{sprcoefffcw}
   S_{fcw} = \rho_l \, \int\limits_{l}^{\infty} \, \left( 
                 \frac{a_3}{z^3} + \frac{a_4}{z^4} \right) \,dz.
\end{equation}   
Noting that, asymptotically for large $l$, the film thickness will
be given by $l \approx - a_4 / a_3$, we arrive at
\begin{equation} \label{ctaasymp}
   1 - \cos{\theta} = - \frac{\rho_l}{6 \gamma_{lv}} \frac{a_3^3}{a_4^2}.
\end{equation}
For partial wetting, we obtain $\theta$ from Young's law, explicitly
calculating $\gamma_{sv}$, $\gamma_{sl}$ and $\gamma_{lv}$ by numerical
minimization, which causes the somewhat noisy results for the contact
angle in this wetting state. Figure 2 shows $\theta$ as a function
of temperature for pentane
on pure water and for hexane on brine ($c_{\rm NaCl} = 2.5$ mol/L) 
\cite{hexonbrine}.
It is seen that not only the transition temperatures for these two systems 
are very similar, but also the contact angles. The discontinuity of $d \theta /
d T$ is clearly visible at $T_{w,1}$.

\begin{figure}
\centerline{\psfig{figure=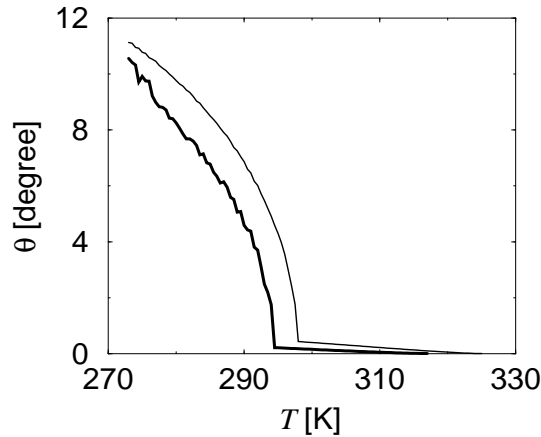,width=7cm,angle=-90}}
\caption{Contact angles of pentane on pure water (thin
line) and of hexane on brine for $c_{\rm NaCl} = 2.5$ mol/L (thick line) as 
a function of temperature. The transition temperatures are $T_{w,1} = 294.5$ K
and $T_{w,c} = 316.7$ K for hexane on brine 
and $T_{w,1} = 298$ K and $T_{w,c} = 325.4$ K for pentane
on water. A value of 1.77 {\AA} has been used for $\delta$ in these cases.}
\end{figure}

In Fig.\ 3, we focus on the behavior of the contact angle in the 
frustrated-complete wetting state; in this state, our data obtained from 
direct numerical minimization are too noisy and, therefore, useless because
the energetical differences between the $cw$ and $fcw$ states are very small. 
The figure shows the analytical results obtained from relation
(\ref{ctaasymp}), which is asymptotically valid in the
case $l \to \infty$. From this 
equation, it can also be read off that $\theta$ is expected to vanish
as $\theta \sim \left( T_{w,c} - T \right)^{3/2}$; this behavior of
the contact angle is, however, only seen very close to $T_{w,c}$.

\begin{figure}
\centerline{\psfig{figure=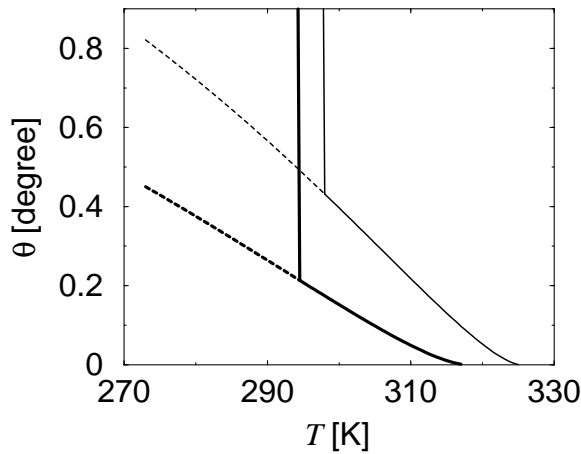,width=7cm,angle=-90}}
\caption{Contact angles near the first-order transition
for pentane on pure water (thin lines) and for hexane on brine ($c_{\rm
NaCl} = 2.5$ mol/L) (thick lines). The solid lines correspond to stable
situations, i.e.\ to partial wetting for $T \le T_{w,1}$ and to 
frustrated-complete wetting for $T_{w,1} \le T \le T_{w,c}$. The dashed
lines indicate metastable extensions of the frustrated-complete wetting
state into the region of $T \le T_{w,1}$, which nevertheless should
be observable experimentally due to a pronounced hysteresis
\protect\cite{shahidzadeh,bertrand,ragil}. }
\end{figure}

In the partial wetting state, upon approach to $T_{w,1}$, the contact
angle (almost, but not quite) vanishes as $\theta \sim \left( T_{w,1} - T
\right)^{1/2}$ as expected for a true first-order {\em wetting}\/
transition. In this case, the (effective) exponent of $1/2$ is seen already 
far from $T_{w,1}$.

\section{Conclusion}
It has been shown that the modified Cahn theory, augmented for 
long-range interactions, which is summarized in Eq.~(\ref{intfactensalt}),
can reproduce the transition temperatures in the sequential wetting of 
$n$-alkanes on (salt) water \cite{weiss3}.
In the present work, we also predict the transition temperatures for pentane 
on brine as a function of salt concentration, which have not been
measured yet. In principle, the theory
allows us to calculate the contact angles as well. In practice, however,
due to limitations of the precision in the numerical calculation, only
the contact angles in the partial wetting state can be determined
reasonably reliably. For the frustrated-complete wetting state, we may
still estimate the asymptotic behavior of $\theta$ from the contributions
of the long-range interactions between substrate and adsorbate alone.

Two of the remaining theoretical tasks are the {\em a priori}\/
calculation of the thickness of the depletion layer $\delta$, which we
find by matching to the experimentally observed transition temperatures
\cite{weiss3},
and the `{\em ab initio}' calculation of the contact energy, which still has 
to be
obtained from measurements of the surface pressure of the entire system
and is, thus, not yet deduced from the properties of the isolated media.

From the experimental point of view, it remains of great interest
to see a realization of a critical endpoint, a point at which the line
of critical wetting transitions is intercepted by the line of first-order 
transitions, which will be first-order {\em wetting}\/ transitions
beyond this point. The discovery of such a system would also demonstrate that 
the parallelism of the two lines found in the system of hexane on brine is 
a coincidence rather than a necessity. Measurements of the contact
angle in the partial and, particularly, in the frustrated-complete wetting
state might be used to test the appropriateness of the present division
of the substrate--adsorbate interactions into short-range and long-range 
forces. The vanishing of the contact angle in the frustrated-complete wetting 
state on approach of $T_{w,c}$ is of primary interested in this respect. 

\section*{Acknowledgements}
We thank Emanuel Bertrand, Daniel Bonn, Harvey Dobbs, Yan Levin, 
Salima Rafa\"{\i} and Ben Widom for fruitful discussions. Our research 
has been supported by the European Community TMR Research Network
`Foam Stability and Wetting' under contract no.\ FMRX-CT98-0171, by
the `Fonds voor Wetenschappelijk Onderzoek (F.W.O.) -- Vlaanderen'
and by the Research Council (`Onderzoeksraad') of the KU Leuven through
junior fellowship F/01/028.


\begin{thebibliography}{99}
\bibitem{hexonbrine} For the cases in which brine is the substrate, $a_3$ 
                     and $a_4$ will have to be 
                     replaced by $a_3 + a_3'$ and $a_4 + a_4'$,
                     respectively, in Eq.~(\protect\ref{ctaasymp}).
\end{thebibliography}
\end{document}